\documentclass[twocolumn]{revtex4}
\usepackage{amssymb}
\usepackage[T1]{fontenc}
\usepackage[latin9]{inputenc}
\usepackage{color}
\usepackage{amsmath}
\usepackage{graphicx}
\usepackage{esint}
\usepackage{bm}

\newcommand{\Tr}{\operatorname{Tr}}

\makeatletter
\@ifundefined{textcolor}{}
{ \definecolor{BLACK}{gray}{0}
 \definecolor{WHITE}{gray}{1}
 \definecolor{RED}{rgb}{1,0,0}
 \definecolor{GREEN}{rgb}{0,1,0}
 \definecolor{BLUE}{rgb}{0,0,1}
 \definecolor{CYAN}{cmyk}{1,0,0,0}
 \definecolor{MAGENTA}{cmyk}{0,1,0,0}
 \definecolor{YELLOW}{cmyk}{0,0,1,0}
 }

\begin{document}

\title{Loophole-free Bell inequality violations cannot disprove local realism}
\author{F. De Zela}

\affiliation{Departamento de Ciencias, Secci\'{o}n
F\'{i}sica \\ Pontificia Universidad Cat\'{o}lica del Per\'{u}, Lima 15088, Peru}




\begin{abstract}
For almost three decades in the twentieth century, the physics community believed that John von Neumann had proved the impossibility of completing quantum mechanics by a local realist, hidden-variables theory. Although Grete Hermann had raised strong objections to von Neumann's proof, she was largely ignored. This situation lasted, until John Bell rediscovered that von Neumann's proof was flawed: a \emph{sufficient} condition for local realism had been taken as a \emph{necessary} one. Bell subsequently established various constraints on hidden-variables theories, in the form of inequalities that can be submitted to experimental test. All performed tests to date have opened some loopholes. The quest to close them motivated great technical achievements and ongoing efforts to improve what has already been reached. There is, however, a rather ironic twist concerning Bell inequalities. On deriving them, Bell also took a sufficient condition for local-realism, as if it were a necessary one. As a consequence, even completely loophole-free Bell inequality violations would not disprove local realism. We argue that Bell inequalities cannot follow from local-realism alone. The proof is given by constructing three local-realist models that entail Bell inequality violations.
\end{abstract}


\maketitle

The challenge of achieving a loophole-free Bell inequality violation has motivated the investment of considerable technical and financial efforts. A few years ago, three different groups almost simultaneously announced to have met that challenge \cite{hensen1,giustina,shalm}. Despite the rather bold and general way in which the announcements were disseminated, the reported achievements were more precisely framed, in each case. It was namely noted that, ``strictly speaking, no Bell experiment can exclude all conceivable local-realist theories'' \cite{hensen1}, that the reported experiments provided just ``the strongest support to date for the viewpoint that local realism is untenable'' \cite{giustina}, and that the announced test \cite{shalm} was loophole free, insofar as only a minimal set of assumptions was made, while ``It is impossible, even in principle, to eliminate a form of these assumptions in any Bell test'' \cite{shalm}. There seems to be a consensus that local-realism has not been completely banned from physics yet \cite{valdenebro,ringbauer,hensen2,santos2022}. For that to be the case, more compelling evidence must be provided. Hence, increasingly sophisticated tests are currently under planning \cite{belenchia2022}. The motivation is not only to provide a definite proof of Bell's claim that ``no theory of nature that obeys locality and realism can reproduce all the predictions of quantum theory`'' \cite{hensen1}. There are also practical motivations. Bell tests could fit some essential needs of quantum information, such as randomness certification and quantum-secure communication. It is thus important to make sure that a loophole-free Bell inequality violation implies that local realism is untenable. We claim that this is not the case. At least not for the Bell inequality that is addressed in the aforementioned tests. This is the Clauser-Horne-Shimony-Holt (CHSH) inequality, or variants thereof (e.g., the Clauser-Horne inequality). 

Let us make precise what is meant by local realism. Locality is understood in its relativistic meaning: physical influences cannot propagate faster than light. Realism means that physical properties exist independently of being measured. That is, measurements reveal preexisting properties. The probabilistic interpretation of quantum mechanics (QM) seemed to be at odds with local realism and this prompted Einstein, Podolsky and Rosen to argue that QM was incomplete \cite{epr}. It was then suggested that hidden variables could deterministically rule the behavior of quantum objects. Their hidden nature would translate into an apparent randomness of quantum phenomena. QM being an incomplete theory, it should be possible to construct a deeper description of physical reality. 

We can identify two main stages regarding the view of hidden-variables theories. During the first stage, a broad community accepted that von Neumann had definitely proved (in 1932) the impossibility of a hidden-variables theory. This acceptance lasted for roughly three decades, even though Grete Hermann had pointed out (in 1935) a flaw in von Neumann's proof \cite{mermin,hermann1935}. Nevertheless, the community's view prevailed over rigorous reasoning. The situation remained basically unchanged, until Bell (in 1966) ``rediscovered the fact that von Neumann's no-hidden variables proof was based on an assumption that can only be described as silly'' \cite{mermin}. However, only in hindsight could von Neumann's assumption be regarded as ``silly''. When regarded in its own context, it appears as a natural and convenient assumption to make. The assumption was the following one. Consider two observables, $\hat{A}$ and $\hat{B}$, whose preexisting values in a hidden-variables theory are $v(\hat{A})$ and $v(\hat{B})$. The preexisting value of $\hat{C}=\hat{A}+\hat{B}$ is assumed to be $v(\hat{C})=v(\hat{A})+v(\hat{B})$. This should hold, even if $\hat{A}$ and $\hat{B}$ do not commute. As pointed out by Mermin \cite{mermin}, von Neumann's assumption that $v(\hat{C})=v(\hat{A})+v(\hat{B})$ holds for each individual system of an ensemble is a sufficient condition for it to hold in the mean, i.e., $\langle \hat{C} \rangle=\langle \hat{A} \rangle + \langle \hat{B} \rangle$, which is required for the hidden-variables theory to be in accordance with QM. While it is clear that von Neumann's assumption is a \emph{sufficient} condition, it is not a \emph{necessary} one. Indeed, one can straightforwardly prove that $v(\hat{C})=v(\hat{A})+v(\hat{B})$ does not generally hold in QM. An example is given by the Pauli matrices $\sigma_x$ and $\sigma_y$ \cite{mermin}. Hence, if von Neumann would have proved that $v(\hat{C})=v(\hat{A})+v(\hat{B})$ \emph{necessarily} holds for any hidden-variables theory, then he would have proved the incompatibility of any such theory with QM. But von Neumann did otherwise and took a sufficient condition as if it were a necessary one, thereby providing a proof which prompted Bell to describe it as ``silly'' \cite{mermin}. Bell then proceeded to develop an alternative approach. His goal was the same as von Neumann's: to show that it is impossible to construct a hidden-variables completion of QM. Bell's approach started the second, current stage regarding the community's view of hidden-variables theories.

As we shall see, what Bell did in fact closely resembles what von Neumann did. Bell's inequality involves correlations $\langle \hat{A}\hat{B}\rangle$ between spin measurements that are performed at two distant sites, $A$ and $B$. At each site, the spin measurements are made along two directions,  $\boldsymbol{\hat{a}}$ and  $\boldsymbol{\hat{a}'}$ at $A$, and 
 $\boldsymbol{\hat{b}}$ and  $\boldsymbol{\hat{b}'}$ at $B$. Bell's assumption is that $\langle \hat{A}\hat{B}\rangle$ is given by an expression whose precise form will be shown below. For now, we just need to point out that the assumed form of $\langle \hat{A}\hat{B}\rangle$ is \emph{sufficient} for it to comply with local realism. On considering various $\langle \hat{A}\hat{B}\rangle$, a quantity $S$ is constructed and proved to be bounded: $|S|\leq 2$. QM predicts the existence of cases for which $|S|>2$. Many experiments have confirmed this prediction, modulo some loopholes. That is, additional assumptions besides local realism were made, each of which opened a loophole. A loophole-free experimental violation of $|S|\leq 2$ allegedly proves that local realism is untenable. As we said, similarly to what happened with von Neumann's assumption, Bell's assumption is a \emph{sufficient} condition for $\langle \hat{A}\hat{B}\rangle$ to comply with local realism. It is hardly a \emph{necessary} condition. Let us mention in advance that Bell's expression for $\langle \hat{A}\hat{B}\rangle$ involves only scalar quantities. However, $\langle \hat{A}\hat{B}\rangle$ is supposed to quantify correlations between measurements that involve three directions: e.g.,  $\boldsymbol{\hat{a}}$,  $\boldsymbol{\hat{b}}$, and a third direction, 
 say $\boldsymbol{\hat{z}}$, that relates to the two-party state which is employed in the measurements. The most general $\langle \hat{A}\hat{B}\rangle$ should depend on vectorial quantities and, possibly, on their scalar products, which account for the relative orientation of $\boldsymbol{\hat{a}}$ and $\boldsymbol{\hat{b}}$. Such a dependence would not be in conflict with locality, because correlations are generally a matter of definition and not a consequence of some cause-effect relationship.

We should also notice that correlations are not directly measured. Correlations are defined per convention and, experimentally, they are constructed from raw measurement data, such as the number of simultaneously detected particles at two distant sites. The conventional character of correlations diminishes the decisive power that an experiment is supposed to have over the validity of theoretical results. Moreover, it should be kept in mind that correlations do not prove a cause-effect relationship, as it is sometimes suggested by referring to nonlocal \emph{effects} between two correlated measurements. 

As already said, we are now in the second stage regarding hidden-variables theories, a stage in which the community's view is that Bell violations do prove wrong local realism, in spite of the above remarks. As it occurred with von Neumann's assumption, the validity of Bell's assumption is taken for granted. However, it is a matter of logic that Bell inequality violations prove local realism wrong only if it is proved that Bell's inequality is a \emph{necessary} consequence of local realism. We could stop here and say that the burden of proof lies with those who claim that Bell's inequality \emph{necessarily} follows from local realism. But we can also go further and show that such a proof cannot be furnished. The reasons are given in what follows.

To begin, let us recall that Bell's correlations are given by
\begin{equation}
  \langle \hat{A}\hat{B} \rangle = \int A_{\boldsymbol{\hat{a}}}(\lambda)B_{\boldsymbol{\hat{b}}}(\lambda)\rho(\lambda)d\lambda. \label{correlation1}
\end{equation}
These correlations correspond to a standard Bell test in which two parties, ``Alice'' and ``Bob'', perform measurements of the Stern-Gerlach (SG) type at two distant sites. The particles on which Alice and Bob perform their measurements are supplied by a source, which produces a pair of spin-1/2 particles and sends one particle to each party. Each SG magnet has a binary output, upward and downward, with respect to its orientation, $\boldsymbol{\hat{a}}$ or $\boldsymbol{\hat{b}}$. Detections at the upward and downward directions are labelled by $\pm 1$, respectively. The quantities entering $\langle \hat{A}\hat{B} \rangle $ have the following meaning: $\hat{A}$ and $\hat{B}$ stand for the observables being measured, in this case the spin components along $\boldsymbol{\hat{a}}$ and $\boldsymbol{\hat{b}}$, respectively; $A_{\boldsymbol{\hat{a}}}(\lambda)=\pm 1$ is the binary output of Alice's SG magnet when it is oriented along $\boldsymbol{\hat{a}}$, this output being determined by the hidden variables $\lambda$, which have a probability distribution $\rho(\lambda)\geq 0$, with $\int \rho(\lambda)d\lambda=1$. $B_{\boldsymbol{\hat{b}}}(\lambda)$ is similarly defined. Bell's assumption is that, in a hidden variables theory, the observable $A_{\boldsymbol{\hat{a}}}B_{\boldsymbol{\hat{b}}}$ has a definite value $(A_{\boldsymbol{\hat{a}}}B_{\boldsymbol{\hat{b}}})(\lambda)$ for each $\lambda$. Moreover, Bell claims \cite{clauser1978} that \emph{a deterministic hidden-variables theory is local if for all $\boldsymbol{\hat{a}}$ and $\boldsymbol{\hat{b}}$ and $\lambda$ we have}
\begin{equation}\label{bell}
(A_{\boldsymbol{\hat{a}}}B_{\boldsymbol{\hat{b}}})(\lambda)=A_{\boldsymbol{\hat{a}}}(\lambda)B_{\boldsymbol{\hat{b}}}(\lambda).
\end{equation}
The expectation value of $A_{\boldsymbol{\hat{a}}}B_{\boldsymbol{\hat{b}}}$ is then given by Eq.~(\ref{correlation1}) and may be taken as a measure for the correlation between the involved observables. Bell's claim is certainly true; but it should be clear that Eq.~(\ref{correlation1}) is a \emph{sufficient} condition for local realism. Bell did not prove that it is a \emph{necessary} one. The same remark holds for what follows.

For simplicity, let us write $A$ and $A^{\prime}$ when referring to the settings $\boldsymbol{\hat{a}}$ and $\boldsymbol{\hat{a}'}$, respectively, and similarly for $\boldsymbol{\hat{b}}$ and $\boldsymbol{\hat{b}'}$. The CHSH measure is defined as
\begin{equation}\label{bm1}
  S=\langle AB \rangle+\langle AB^{\prime} \rangle +\langle A^{\prime}B \rangle -\langle A^{\prime}B^{\prime} \rangle.
\end{equation}
The structure of $S$ comes from considering the function
\begin{equation}\label{f}
f(\lambda)=(A(\lambda)+A^{\prime}(\lambda))B(\lambda)+(A(\lambda)-A^{\prime}(\lambda))B^{\prime}(\lambda),
\end{equation}
which is the sum of two terms, one of which is $\pm 2$ and the other zero for each $\lambda$, whereupon $f(\lambda)=\pm 2$. From this and
$|\int f(\lambda)\rho(\lambda)d\lambda| \leq \int |f(\lambda)\rho(\lambda)|d\lambda=2$, one gets
\begin{equation}\label{s}
  |S|\leq 2,
\end{equation}
which is the Bell-CHSH inequality.

Experiments designed to violate inequality (\ref{s}) use entangled states, viz., states that cannot be written as a product of two terms, one related to $A$ and the other to $B$. These experiments have much more to do with entanglement than with local realism \cite{gisin1991}, as we shall see. 

It is clear that inequality (\ref{s}) cannot be submitted directly to experimental test, because the outputs of SG-type experiments are not correlations $\langle AB \rangle_{\text{exp}}$, but number of registered particles at each detector. One records the number of coincident detections during some fixed interval of time. There are two detectors at each site, one for each output of the SG magnet.  
Let us denote by $N_{\boldsymbol{\hat{a}},\boldsymbol{\hat{b}}}(\alpha,\beta)$, with $\alpha=\pm 1$ and $\beta=\pm 1$, the number of coincident detections at SG magnets oriented along $\boldsymbol{\hat{a}}$ and $\boldsymbol{\hat{b}}$, with detections occurring at the upward ($+1$) or downward ($-1$) outputs. Having recorded a sufficiently large number of coincidences, one can assess the corresponding probabilities for simultaneous detections: $P_{\boldsymbol{\hat{a}},\boldsymbol{\hat{b}}}^{(\text{exp})}(\alpha,\beta)=N_{\boldsymbol{\hat{a}},\boldsymbol{\hat{b}}}(\alpha,\beta)/N_{\text{tot}}$, with $N_{\text{tot}}=\sum_{\alpha,\beta}N_{\boldsymbol{\hat{a}},\boldsymbol{\hat{b}}}(\alpha,\beta)$. The experimental correlations are constructed as
\begin{equation}\label{expcor}
  \langle \hat{A} \hat{B} \rangle_{\text{exp}}=\frac{1}{N_{\text{tot}}}\sum_{\alpha,\beta} \alpha \beta N_{\boldsymbol{\hat{a}},\boldsymbol{\hat{b}}}(\alpha,\beta)=\sum_{\alpha,\beta} \alpha \beta P_{\boldsymbol{\hat{a}},\boldsymbol{\hat{b}}}^{(\text{exp})}(\alpha,\beta),
\end{equation}
and are then used to obtain the experimental CHSH measure $S_{\text{exp}}$, as per Eq.~(\ref{bm1}). There are numerous experiments in which $|S_{\text{exp}}|>2$ have been observed. Even if all their loopholes would have been closed, one cannot claim that these experiments falsify local realism. As already said, for this to be the case, one should prove that Eq.~(\ref{correlation1}) \emph{necessarily} follows from  local realism. Alternatively, one can address probabilities. These are conceptually more fundamental than expectation values or correlations. The connection between probabilities and measurement results, such as $A_{\boldsymbol{\hat{a}}}$ and $B_{\boldsymbol{\hat{b}}}$, is given by
\begin{equation}\label{connection}
 P_{\boldsymbol{\hat{a}},\boldsymbol{\hat{b}}}(\alpha,\beta)=\int \left(\frac{1+\alpha A_{\boldsymbol{\hat{a}}}(\lambda)}{2}\right)\left(\frac{1+\beta B_{\boldsymbol{\hat{b}}}(\lambda)}{2}\right)\rho(\lambda) d\lambda,
\end{equation}
and similarly for the cases involving $\boldsymbol{\hat{a}'}$ and $\boldsymbol{\hat{b}'}$. From Eq.~(\ref{connection}), one can straightforwardly obtain
\begin{equation}\label{expcor2}
\langle \hat{A}\hat{B} \rangle=\sum_{\alpha,\beta} \alpha \beta P_{\boldsymbol{\hat{a}},\boldsymbol{\hat{b}}}(\alpha,\beta) = \int A_{\boldsymbol{\hat{a}}}(\lambda)B_{\boldsymbol{\hat{b}}}(\lambda)\rho(\lambda)d\lambda,
\end{equation}
in accordance with Eq.~(\ref{correlation1}).
Equation (\ref{connection}) can be written as follows:
\begin{eqnarray}
 P_{\boldsymbol{\hat{a}},\boldsymbol{\hat{b}}}(\alpha,\beta)&\equiv&\int p(\alpha,\beta|\boldsymbol{\hat{a}},\boldsymbol{\hat{b}},\lambda)\rho(\lambda) d\lambda \label{connection2a}\\
 &=&\int p(\alpha|\boldsymbol{\hat{a}},\lambda)\,p(\beta|\boldsymbol{\hat{b}},\lambda)\rho(\lambda) d\lambda. \label{connection2b}
\end{eqnarray}
Here, $p(\alpha,\beta|\boldsymbol{\hat{a}},\boldsymbol{\hat{b}},\lambda)$ is the conditional probability distribution for Alice and Bob obtaining $\alpha$ and $\beta$ as results of their spin measurements, provided these measurements have been performed with their SG magnets oriented along
$\boldsymbol{\hat{a}}$ and $\boldsymbol{\hat{b}}$, respectively, while the hidden variables took on the value $\lambda$. Likewise,
$p(\alpha|\boldsymbol{\hat{a}},\lambda)$ denotes the probability distribution for Alice's spin measurement to yield $\alpha$, if her SG magnet was oriented along $\boldsymbol{\hat{a}}$ and the hidden variables took on the value $\lambda$. Similarly for $p(\beta|\boldsymbol{\hat{b}},\lambda)$.  

As can be seen, the probability distribution for $\alpha$ and $\beta$ factorizes:
\begin{equation}\label{factor}
 p(\alpha,\beta|\boldsymbol{\hat{a}},\boldsymbol{\hat{b}},\lambda)=
 p(\alpha|\boldsymbol{\hat{a}},\lambda)\,p(\beta|\boldsymbol{\hat{b}},\lambda).
\end{equation}
The above decomposition ``represents a precise condition for locality in the context of Bell experiments''
\cite{brunner2014}. We may rephrase this statement by saying that the decomposition in Eq.~(\ref{factor}) is a \emph{sufficient} condition for local realism. Although it is not a \emph{necessary} condition, it seems to have been generally taken as such. It was  perhaps too appealing that, from this simple relationship, an inequality such as $|S|\leq 2$ could be straightforwardly derived and used to submit local realism to experimental test.

A salient feature of Eq.~(\ref{factor}) is the inclusion of only scalar quantities, even though it refers to measurements for which directions play an essential role. The vectors $\boldsymbol{\hat{a}}$ and $\boldsymbol{\hat{b}}$ appear only as labels. 
As appealing as it is, Eq.~(\ref{factor}) does not exhaust all the possibilities we have to define a probability measure. Indeed, we can proceed as follows. Let us consider an inner product vector space $V$ and a continuous function $f$ that is orthogonally additive and maps vectors $v \in V$ to the reals, i.e., $f:V \rightarrow \mathbb{R}$. We denote the inner product by $v \cdot v^{\prime}$. A function $f$ is orthogonally additive, if $f(v+v^{\prime})=f(v)+f(v^{\prime})$, whenever $ v \cdot v^{\prime}=0$. Gudder's theorem \cite{Gudder} states that if $f:V \rightarrow \mathbb{R}$ is orthogonally additive and continuous, then it is of the form
\begin{equation}\label{gudder}
  f(v)=c (v \cdot v)+ k\cdot v,
\end{equation}
where $c \in \mathbb{R}$ and $k\in V$.

As we are interested in Bell tests of the CHSH type, we may focus on two-state systems. These are represented by a (generally unnormalized) ket, e.g., $|\psi\rangle=\alpha |\uparrow\rangle + \beta |\downarrow\rangle$. Correspondingly, we focus on the space $V_4$, with elements $(v_0,v_1,v_2,v_3)\equiv (v_0,\mathbf{v})$. These vectors can be put in one-to-one correspondence with $|\psi \rangle \langle \psi |$, via
\begin{equation}
|\psi \rangle \langle \psi | =\frac{1}{2}
\sum_{\mu=0}^{3} v_{\mu} \sigma_{\mu}, \label{roy}
\end{equation}
where $\sigma_{0}$ is the identity and $\sigma_{\mu=1,2,3}$ the Pauli matrices.

We can now define a measure $f_{\phi}$ relative to any reference vector
$\boldsymbol{v}_{\phi}\equiv (1,\mathbf{\hat{n}}_{\phi})$, with $\mathbf{\hat{n}}_{\phi}$ being a three-dimensional unit vector. We take $f_{\phi}$ to be orthogonally additive and continuous. Continuity guarantees that very small changes of preexisting values are reflected in similar changes of measured values. The requirement that $f_{\phi}$ is orthogonally additive is just an instance of a general requirement that is put on measures. These are generally defined as a map $m$ from elements of an appropriate set (a so-called $\sigma$-algebra) to, e.g., the real numbers. The map $m$ is required to satisfy $m(A \cup B)=m(A)+m(B)$, whenever $A\cap B =\emptyset$.

Additionally, we put on $f_{\phi}$ the following requirements:

\noindent 1) $f_{\phi}(\boldsymbol{v}_{\phi})=1$.

\noindent 2) $f_{\phi}(\boldsymbol{v}_{\phi _{\perp}})=f_{\phi}(\boldsymbol{v}_{\phi^{\prime}_{\perp}})=0$ for vectors
$\boldsymbol{v}_{\phi_{\perp}}\equiv (1,-\mathbf{\hat{n}}_{\phi})$ and
$\boldsymbol{v}_{\phi^{\prime}_{\perp}}\equiv (-1,\mathbf{\hat{n}}_{\phi})$ which are orthogonal to $\boldsymbol{v}_{\phi}$.

\noindent 3) $f_{\phi}(\boldsymbol{v}_{\psi}) \in \left[0,1\right]$ for any vector $\boldsymbol{v}_{\psi}=(1,\mathbf{\hat{n}}_{\psi})$. 

The first requirement means that $f_{\phi}$ fits exactly one time into itself, something that should happen with any sensible measure. Indeed, measurements essentially consist on counting how many times a given unit fits into what is being measured.
The second requirement consistently complements the first one. The third requirement allows us to use $f_{\phi}$ as a probability measure. Of course, all this is a matter of convention. There is nothing inherently ``classical'' or ``quantal'' in it.

From Gudder's theorem and the above requirements, it follows that \cite{fdz4}
\begin{equation}\label{final}
  f_{\phi}(\boldsymbol{v}_{\psi})=\frac{1}{2}(1,\mathbf{\hat{n}}_{\phi})\cdot(1,\mathbf{\hat{n}}_{\psi})=\frac{1}{2}\left( 1+\mathbf{\hat{n}}_{\phi}\cdot \mathbf{\hat{n}}_{\psi}\right).
\end{equation}

The probability measure $f_{\phi}(\boldsymbol{v}_{\psi})$ is in fact closely related to what is called a ``quantum'' probability. The latter refers to Born's rule for the probability of measuring the state $|\psi\rangle$ on a system that has been prepared in state $|\phi\rangle$. This probability is given by $|\langle \psi|\phi\rangle|^2$, provided the two states are normalized. In terms of the projectors $\widehat{\Pi}_{\phi}= |\phi\rangle\langle\phi|$ and $\widehat{\Pi}_{\psi}=|\psi\rangle\langle\psi|$, we can write Born's rule in the form
\begin{equation}\label{born}
|\langle \psi|\phi\rangle|^2=\Tr\left(\widehat{\Pi}_{\phi}\widehat{\Pi}_{\psi}\right)=\frac{1}{2}\left( 1+\mathbf{\hat{n}}_{\phi}\cdot \mathbf{\hat{n}}_{\psi}\right),
\end{equation}
where we have used $\widehat{\Pi}_{\phi}=(1/2)(\sigma_0+\mathbf{\hat{n}}_{\phi} \cdot \boldsymbol{\sigma})$ and similarly for $\widehat{\Pi}_{\psi}$. We see therefore that
\begin{equation}\label{final2}
   f_{\phi}(\boldsymbol{v}_{\psi})=\frac{1}{2}\left( 1+\mathbf{\hat{n}}_{\phi}\cdot \mathbf{\hat{n}}_{\psi}\right)=|\langle \psi|\phi\rangle|^2.
\end{equation}
This shows, first, that ``quantum'' probabilities $|\langle \psi|\phi\rangle|^2$ fit well in the general scheme we have given for probability measures. What is essential for this scheme is the linear vector space structure on which it is based. Second, the probabilities we have used in the context of Bell's inequality are too narrowly framed. They are limited by their dependence on scalar quantities alone, and by having a structure which does not incorporate entangled quantities. Let us see how this limitations can be removed. Our framework will continue to be just that of a linear vector space. 

We want to address two-particle systems. Each particle is supposed to carry a two-state degree of freedom (DOF), such as spin-1/2, polarization, two-way paths, etc. For the description of a single particle having such a DOF, we can use a vector space $V_4$ with elements $\boldsymbol{v}=\left( v_0,v_1,v_2,v_3 \right)$ and Euclidean inner product
$\langle \boldsymbol{v}, \boldsymbol{v}^{\prime}\rangle=\sum_{\mu=0}^{3} v_{\mu}v_{\mu}^{\prime}$. When addressing two particles, $A$ and $B$, we use the tensor product space $V^{AB}=V_{4}^{A}\otimes V_{4}^{B}$ with orthonormal basis
$\left\{\boldsymbol{\hat{e}}_{\mu}^{A}\otimes \boldsymbol{\hat{e}}_{\nu}^{B}\right\}$ ($\mu, \nu \in \{0,\ldots,3\}$). We define the inner product
$\langle \boldsymbol{v}_A \otimes \boldsymbol{w}_B,\boldsymbol{v}^{\prime}_A \otimes \boldsymbol{w}^{\prime}_B \rangle=
\langle \boldsymbol{v}_A,\boldsymbol{v}^{\prime}_A\rangle \langle \boldsymbol{w}_B ,\boldsymbol{w}^{\prime}_B \rangle$, and extend it to all $V^{AB}$ by linearity. Gudder's theorem let us connect inner-product measures with correlations. With this in mind, we represent Alice's and Bob's observables by the following vectors:
\begin{eqnarray}\label{table3}
\boldsymbol{\mathcal{A}}&=&a_1\boldsymbol{\hat{e}}_{1}^{A} +a_2\boldsymbol{\hat{e}}_{2}^{A} +a_3\boldsymbol{\hat{e}}_{3}^{A}, \\
\boldsymbol{\mathcal{B}}&=&b_1\boldsymbol{\hat{e}}_{1}^{B} +b_2\boldsymbol{\hat{e}}_{2}^{B} +b_3\boldsymbol{\hat{e}}_{3}^{B}, 
\end{eqnarray}
which correspond to the operators $\hat{A}=\boldsymbol{\hat{a}} \cdot \boldsymbol{\sigma}^{A}$ and $\hat{B}=\boldsymbol{\hat{b}} \cdot \boldsymbol{\sigma}^{B}$, respectively, of the quantum formalism. Here, $\boldsymbol{\sigma}$ is the triple of Pauli matrices.

The source delivers two-particle states, which in the present framework are represented by vectors $\boldsymbol{\phi}_{AB} \in V^{AB}$. Correlations between Alice's and Bob's measurements along directions $\boldsymbol{\hat{a}}$ and $\boldsymbol{\hat{b}}$, respectively, are given by the inner product
$\langle\boldsymbol{\mathcal{A}}\otimes\boldsymbol{\mathcal{B}},\boldsymbol{\phi}_{AB} \rangle$. This last prescription is, up to normalization, just an application of Gudder's theorem to the case of the tensor-product space $V^{AB}$ \cite{fdz5}. In Bell tests, one usually employs one of the four Bell states. In the present formalism, these states are prescribed to be given by 
\begin{subequations}
\begin{align}\label{a1}
\boldsymbol{\phi}_{AB}^{+}&=&\frac{1}{2} \left(\boldsymbol{\hat{e}}_{0}^{A}\otimes\boldsymbol{\hat{e}}_{0}^{B}+ \boldsymbol{\hat{e}}_{1}^{A}\otimes\boldsymbol{\hat{e}}_{1}^{B}-\boldsymbol{\hat{e}}_{2}^{A}\otimes\boldsymbol{\hat{e}}_{2}^{B}+
\boldsymbol{\hat{e}}_{3}^{A}\otimes\boldsymbol{\hat{e}}_{3}^{B}\right),\\
\boldsymbol{\phi}_{AB}^{-}&=&  \frac{1}{2}\left(\boldsymbol{\hat{e}}_{0}^{A}\otimes\boldsymbol{\hat{e}}_{0}^{B} -\boldsymbol{\hat{e}}_{1}^{A}\otimes\boldsymbol{\hat{e}}_{1}^{B}+\boldsymbol{\hat{e}}_{2}^{A}\otimes\boldsymbol{\hat{e}}_{2}^{B}+
\boldsymbol{\hat{e}}_{3}^{A}\otimes\boldsymbol{\hat{e}}_{3}^{B}\right), \\
\boldsymbol{\psi}_{AB}^{+}&=& \frac{1}{2} \left(\boldsymbol{\hat{e}}_{0}^{A}\otimes\boldsymbol{\hat{e}}_{0}^{B}+ \boldsymbol{\hat{e}}_{1}^{A}\otimes\boldsymbol{\hat{e}}_{1}^{B}+\boldsymbol{\hat{e}}_{2}^{A}\otimes\boldsymbol{\hat{e}}_{2}^{B}-
\boldsymbol{\hat{e}}_{3}^{A}\otimes\boldsymbol{\hat{e}}_{3}^{B}\right),\\
\boldsymbol{\psi}_{AB}^{-}&=&\frac{1}{2} \left(\boldsymbol{\hat{e}}_{0}^{A}\otimes\boldsymbol{\hat{e}}_{0}^{B} -\boldsymbol{\hat{e}}_{1}^{A}\otimes\boldsymbol{\hat{e}}_{1}^{B}-\boldsymbol{\hat{e}}_{2}^{A}\otimes\boldsymbol{\hat{e}}_{2}^{B}-
\boldsymbol{\hat{e}}_{3}^{A}\otimes\boldsymbol{\hat{e}}_{3}^{B}\right).
\end{align}
\end{subequations}
The above vectors are not factorable, i.e., they cannot be written in the form
$\boldsymbol{\hat{v}}^{A}\otimes\boldsymbol{\hat{w}}^{B}$. So, they fit the definition of ``entangled states''. This definition applies for tensor-product spaces, irrespective of their employment in a quantum or classical context \cite{spreeuw,eberly1,qian,borges,kagalwala,stoklasa,mclaren,sandeau}. From the definition of the inner product in $V^{AB}$, it readily follows that
\begin{subequations}\label{qcorr0}
\begin{eqnarray} \label{qcorr1}
\langle \boldsymbol{\mathcal{A}}\otimes \boldsymbol{\mathcal{B}},\boldsymbol{\phi}_{AB}^{+} \rangle &=& \frac{1}{2}(a_1 b_1-a_2 b_2 +a_3 b_3) \label{b1}, \\
 \langle \boldsymbol{\mathcal{A}}\otimes \boldsymbol{\mathcal{B}},\boldsymbol{\phi}_{AB}^{-} \rangle &=& \frac{1}{2} (-a_1 b_1 +a_2 b_2+a_3 b_3) \label{b2}, \\
\langle \boldsymbol{\mathcal{A}}\otimes \boldsymbol{\mathcal{B}},\boldsymbol{\psi}_{AB}^{+} \rangle &=& \frac{1}{2}(a_1 b_1+a_2 b_2-a_3 b_3) \label{b3}, \\
\langle \boldsymbol{\mathcal{A}}\otimes \boldsymbol{\mathcal{B}},\boldsymbol{\psi}_{AB}^{-} \rangle&=& \frac{1}{2}(-a_1 b_1-a_2 b_2-a_3 b_3) \label{b4} .
\end{eqnarray}
\end{subequations}
On absorbing the factor $1/2$ in the definition of the correlations, we obtain the same results that follow from the quantum formalism. These are given by, e.g., 
\begin{equation}\label{mv}
\langle \Phi^{+}_{AB}| (\boldsymbol{\hat{a}} \cdot \boldsymbol{\sigma}^{A})\otimes(\boldsymbol{\hat{b}} \cdot \boldsymbol{\sigma}^{B})|\Phi^{+}_{AB}\rangle =a_1 b_1-a_2 b_2 +a_3 b_3 ,
\end{equation}
and similarly for all Bell states:
\begin{eqnarray}
 |\Phi^{\pm}_{AB}\rangle &=& \frac{1}{\sqrt{2}}(|\uparrow\rangle|\uparrow\rangle\pm|\downarrow\rangle|\downarrow\rangle), \\
 |\Psi^{\pm}_{AB}\rangle &=& \frac{1}{\sqrt{2}} (|\uparrow\rangle|\downarrow\rangle\pm|\downarrow\rangle|\uparrow\rangle).
\end{eqnarray}

On applying Gudder's theorem, we get the probabilities for simultaneous detections with results $\alpha$ and $\beta$ at SG magnets oriented along $\boldsymbol{\hat{a}}$ and $\boldsymbol{\hat{b}}$:
\begin{align}
  P_{\boldsymbol{\hat{a}},\boldsymbol{\hat{b}}}^{\boldsymbol{\phi}_{AB}}(\alpha,\beta)&=&\frac{1}{4}\langle (1,\alpha\boldsymbol{\hat{a}})\otimes (1,\beta\boldsymbol{\hat{b}}),\boldsymbol{\phi}_{AB} \rangle \\
  &=&
  \frac{1}{4}\left[1+\alpha \beta \, \langle \boldsymbol{\mathcal{A}}\otimes \boldsymbol{\mathcal{B}},\boldsymbol{\phi}_{AB} \rangle \right].
\end{align}
Here, $\boldsymbol{\phi}_{AB}$ stands for any Bell state in which the system was prepared. Equations (\ref{qcorr0}) show that the correlations $\langle \boldsymbol{\mathcal{A}}\otimes \boldsymbol{\mathcal{B}},\boldsymbol{\phi}_{AB} \rangle$ are given by a scalar product of two vectors that differ from
$\boldsymbol{\hat{a}}$ or $\boldsymbol{\hat{b}}$ only in the sign of one or more components. The algebraic structure of $P_{\boldsymbol{\hat{a}},\boldsymbol{\hat{b}}}^{\boldsymbol{\phi}_{AB}}(\alpha,\beta)$ thus markedly differs from that of 
$P_{\boldsymbol{\hat{a}},\boldsymbol{\hat{b}}}(\alpha,\beta)$ (see Eq.~(\ref{connection})).

It should be clear that nothing in the above formulation brings it into conflict with local realism, and yet it is in accord with Bell inequality violations. The above formulation thus shows that $P_{\boldsymbol{\hat{a}},\boldsymbol{\hat{b}}}(\alpha,\beta)$ was too narrowly framed. It does not have the most general structure that is compatible with local realism. The correlations $\langle \hat{A}\hat{B} \rangle$ that follow from $P_{\boldsymbol{\hat{a}},\boldsymbol{\hat{b}}}(\alpha,\beta)$ (see Eq.~(\ref{expcor2})) have a structure in which entanglement is, at most, only nominally referred to, via the hidden variables $\lambda$. These are supposed to define the state in which the two-particle system was prepared. However, one cannot explicitly distinguish an entangled state from a factorable one, because the very structure of $\langle \hat{A}\hat{B} \rangle$ does not allow us to make such a distinction. 

In our description, we have not included hidden variables. Though this was irrelevant to our purposes, we can extend our description to include them. 
Indeed, we can generalize what we did before, by considering a continuous, orthonormal basis in $V^{A}$, i.e.,
\begin{equation}\label{innerprodbasis}
\langle \boldsymbol{\hat{e}}_{\mu}^{A}(\lambda), \boldsymbol{\hat{e}}_{\nu}^{A}(\lambda^{\prime})\rangle=\delta_{\mu\nu}\delta(\lambda-\lambda^{\prime}),
\end{equation}
and similarly in $V^{B}$. We then proceed as before and define the inner product
\begin{equation}\label{innerprodbasis2}
\begin{aligned}
\langle \boldsymbol{\hat{e}}_{\mu}^{A}(\lambda) \otimes \boldsymbol{\hat{e}}_{\sigma}^{B}(\lambda),\boldsymbol{\hat{e}}_{\nu}^{A}(\lambda^{\prime})\otimes \boldsymbol{\hat{e}}_{\tau}^{B}(\lambda^{\prime}) \rangle =\\
=
\langle \boldsymbol{\hat{e}}_{\mu}^{A}(\lambda),\boldsymbol{\hat{e}}_{\nu}^{A}(\lambda^{\prime})\rangle \, \langle \boldsymbol{\hat{e}}_{\sigma}^{B}(\lambda) ,\boldsymbol{\hat{e}}_{\tau}^{B}(\lambda^{\prime}) \rangle,
\end{aligned}
\end{equation}
and extend it to all $V^{AB}$ by linearity.
Alice's and Bob's observables also depend on hidden variables. However, they can freely choose the orientation of their SG magnets, thereby fixing the respective hidden variables:
\begin{eqnarray}
\boldsymbol{\mathcal{A}}(\lambda_A)&=&a_1\boldsymbol{\hat{e}}_{1}^{A}(\lambda_A) +a_2\boldsymbol{\hat{e}}_{2}^{A}(\lambda_A) +a_3\boldsymbol{\hat{e}}_{3}^{A}(\lambda_A), \label{table3a}\\
\boldsymbol{\mathcal{B}}(\lambda_B)&=&b_1\boldsymbol{\hat{e}}_{1}^{B}(\lambda_B) +b_2\boldsymbol{\hat{e}}_{2}^{B}(\lambda_B) +b_3\boldsymbol{\hat{e}}_{3}^{B}(\lambda_B). \label{table3b}
\end{eqnarray}
We assume that $\lambda_A \approx\lambda_B$, which reflects that Alice and Bob use SG-magnets that are supposed to be identically constructed, from a macroscopic point of view. We admit some randomness, though, so that the hidden variables at Alice's and Bob's sites are governed by a probability distribution $\rho_M(\lambda)$, which is sharply peaked at some value $\lambda_M$, so that
$\lambda_A \approx\lambda_B\approx \lambda_M$.

As for the source, it is set to deliver, e.g., the Bell state 
\begin{equation}\label{phi}
\begin{aligned}
  \boldsymbol{\phi}_{AB}^{+}=\frac{1}{\sqrt{N}}\int_{-\infty}^{\infty} \left[\boldsymbol{\hat{e}}_{0}^{A}(\lambda)\boldsymbol{\hat{e}}_{0}^{B}(\lambda) +\boldsymbol{\hat{e}}_{1}^{A}(\lambda)\boldsymbol{\hat{e}}_{1}^{B}(\lambda) \right.\\ 
  \left. -\,\boldsymbol{\hat{e}}_{2}^{A}(\lambda)\boldsymbol{\hat{e}}_{2}^{B}(\lambda)+
\boldsymbol{\hat{e}}_{3}^{A}(\lambda)\boldsymbol{\hat{e}}_{3}^{B}(\lambda)\right]\rho_S(\lambda)d\lambda,
\end{aligned}
\end{equation}
where, for brevity, we have suppressed the tensor-product symbol, and the normalization factor reads $N=4\int_{-\infty}^{\infty} \rho_S^2(\lambda)d\lambda$. The probability distribution at the source, $\rho_S(\lambda)$, is also supposed to be sharply peaked at $\lambda=\lambda_S$. We may further assume that $\lambda_S \approx \lambda_M$. This reflects that the orientations of the SG-magnets and that of the apparatuses used to produce $\boldsymbol{\phi}_{AB}^{+}$ are subjected to similar conditions. 

On observing that, e.g.,
\begin{equation}\label{dirac}
\begin{aligned}
 & \langle \boldsymbol{\hat{e}}_{\mu}^{A}(\lambda_A)\boldsymbol{\hat{e}}_{\nu}^{B}(\lambda_B),\int_{-\infty}^{\infty}\boldsymbol{\hat{e}}_{\sigma}^{A}(\lambda)\boldsymbol{\hat{e}}_{\tau}^{B}(\lambda)\rho_S(\lambda)d\lambda\rangle =\\
 & =\delta_{\mu\sigma}\delta_{\nu\tau}\int_{-\infty}^{\infty}\delta(\lambda_A-\lambda)\delta(\lambda_B-\lambda)\rho_S(\lambda)d\lambda=\\
 &=\delta_{\mu\sigma}\delta_{\nu\tau}\delta(\lambda_A-\lambda_B)\rho_S(\lambda_A),
  \end{aligned}
\end{equation}
we obtain
\begin{equation}\label{correlationf}
\langle \boldsymbol{\mathcal{A}}\otimes \boldsymbol{\mathcal{B}},\boldsymbol{\phi}_{AB}^{+} \rangle =\frac{\delta(\lambda_A-\lambda_B)\rho_S(\lambda_A)}{\sqrt{N}} (a_1 b_1-a_2 b_2 +a_3 b_3). 
\end{equation}
We can now take, e.g., Alice's setting $\lambda_A$ as a reference, and let Bob's setting $\lambda_B$ vary, so that $\rho_M(\lambda_B)d\lambda_B$ is the number of cases in $[\lambda_B,\lambda_B +d\lambda_B]$, for which $\lambda_B\approx\lambda_A$. We thus define the following correlation
\begin{equation}\label{correlationf2}
\langle \boldsymbol{\mathcal{A}}\otimes \boldsymbol{\mathcal{B}},\boldsymbol{\phi}_{AB}^{+} \rangle_T =
\int \langle \boldsymbol{\mathcal{A}}\otimes \boldsymbol{\mathcal{B}},\boldsymbol{\phi}_{AB}^{+} \rangle \rho_M(\lambda_B) d\lambda_B, 
\end{equation}
as the one which can be compared with experimental outputs. From Eqs.~(\ref{correlationf}) and (\ref{correlationf2}), we get
\begin{eqnarray}\label{correlationf3}
\langle \boldsymbol{\mathcal{A}}\otimes \boldsymbol{\mathcal{B}},\boldsymbol{\phi}_{AB}^{+} \rangle_T &=&\frac{\rho_S(\lambda_A)\rho_M(\lambda_A)}{\sqrt{N}} (a_1 b_1-a_2 b_2 +a_3 b_3) \nonumber \\
&\approx& \kappa_{SM} (a_1 b_1-a_2 b_2 +a_3 b_3), 
\end{eqnarray}
where we have set
\begin{equation}\label{kappa}
  \kappa_{SM}=\frac{\rho_S(\lambda_S)\rho_M(\lambda_M)}{\sqrt{N}}.
\end{equation}
This factor can be absorbed by including it in the definitions of $\boldsymbol{\mathcal{A}}(\lambda_A)$ and $\boldsymbol{\mathcal{B}}(\lambda_B)$ (see Eqs.~(\ref{table3a}) and (\ref{table3b})), so that we have again the results given by QM.  

Another way to include hidden variables in our description is as follows. We take just one hidden variable $\lambda$, which has the meaning of being the probability for the source to deliver a Bell state, say,
\begin{equation}\label{bellphi}
\boldsymbol{\phi}_{AB}^{+}=\frac{1}{2} \left(\boldsymbol{\hat{e}}_{0}^{A}\otimes\boldsymbol{\hat{e}}_{0}^{B}+ \boldsymbol{\hat{e}}_{1}^{A}\otimes\boldsymbol{\hat{e}}_{1}^{B}-\boldsymbol{\hat{e}}_{2}^{A}\otimes\boldsymbol{\hat{e}}_{2}^{B}+
\boldsymbol{\hat{e}}_{3}^{A}\otimes\boldsymbol{\hat{e}}_{3}^{B}\right).
\end{equation}
Otherwise, the source delivers a completely random state:
\begin{equation}\label{random}
\boldsymbol{I}_{AB}=\boldsymbol{\hat{e}}_{0}^{A}\otimes\boldsymbol{\hat{e}}_{0}^{B}.
\end{equation}
Hence, the source delivers the Werner-type states
\begin{equation}\label{bellphi2}
\boldsymbol{\Phi}_{AB}^{+}=\lambda \, \boldsymbol{\phi}_{AB}^{+}+(1-\lambda)\boldsymbol{I}_{AB}.
\end{equation}
In this case, we obtain the correlation
\begin{equation}\label{sec}
  \langle \boldsymbol{\mathcal{A}}\otimes \boldsymbol{\mathcal{B}},\boldsymbol{\Phi}_{AB}^{+} \rangle =\lambda (a_1 b_1-a_2 b_2 +a_3 b_3). 
\end{equation}
Here again, the factor $\lambda$ can be absorbed by including it in the definitions of the observables $\boldsymbol{\mathcal{A}}$ and $\boldsymbol{\mathcal{B}}$. 

It is instructive to consider a third model, in which we assume that the vectors entering the expression for $\boldsymbol{\phi}_{AB}^{+}$, see Eq.~(\ref{bellphi}), are randomly produced according to
\begin{eqnarray}
  \boldsymbol{\hat{e}}_{\mu}^{A}&=&\frac{1}{\sqrt{N_A}}\int_{-\infty}^{\infty}\boldsymbol{\hat{e}}_{\mu}^{A}(\lambda)\rho_A(\lambda)d\lambda, \label{m3a}\\
  \boldsymbol{\hat{e}}_{\mu}^{B}&=&\frac{1}{\sqrt{N_B}}\int_{-\infty}^{\infty}\boldsymbol{\hat{e}}_{\mu}^{B}(\lambda^{\prime})\rho_B(\lambda^{\prime})d\lambda^{\prime},\label{m3b}
\end{eqnarray}
where $N_k=\int_{-\infty}^{\infty}\rho_k^2(\lambda)d\lambda$, $k=A,B$.
In contrast to Eq.~(\ref{phi}), here we do not assume that a single hidden variable $\lambda$ is shared by the vectors entering $\boldsymbol{\phi}_{AB}^{+}$. That is, in the present case, $\boldsymbol{\hat{e}}_{\mu}^{A}(\lambda)$ and $\boldsymbol{\hat{e}}_{\mu}^{B}(\lambda^{\prime})$ have independent random fluctuations when produced at the source. There is no ``shared randomness'' in this case. Nevertheless, we readily obtain
\begin{equation}\label{correlationf4}
\langle \boldsymbol{\mathcal{A}}\otimes \boldsymbol{\mathcal{B}},\boldsymbol{\phi}_{AB}^{+} \rangle =\frac{\rho_A(\lambda_A)\rho_B(\lambda_B)}{\sqrt{N}} (a_1 b_1-a_2 b_2 +a_3 b_3). 
\end{equation}
This case shows that the entangled nature of the state can be accounted for in a local realistic model without assuming ``shared randomness''. The latter can be incorporated, as in Bell's model, but as long as entanglement is absent, no Bell violation should be expected.

The above examples show that Bell-type correlations, i.e., those given by Eq.~(\ref{correlation1}), while being in accord with local realism, constitute a very limited family. The reasons for which Eq.~(\ref{correlation1}) complies with local realism also apply, term by term, to the expressions leading to Eqs.~(\ref{correlationf}), (\ref{correlationf2}), (\ref{sec}) and (\ref{correlationf4}). This shows that Bell-type correlations, and the inequality that these correlations imply, do not derive from local realism alone. 

The formalism we have developed has two essential features. First, a probability measure that arises from a linear vector space structure when we put some requirements on a general measure, so as to make it appropriate for being used as a probability measure. Second, the involvement of a tensor-product structure, which allows us to deal with entangled states. There is nothing ``quantum'' in all this. It applies at both the quantal level and the classical level, whatever the distinction between these levels might be.

As for the experimental Bell inequality violations, they involve three types of correlations: Bell-type correlations $\langle \hat{A}\hat{B} \rangle_{\text{Bell}}$, given by Eq.~(\ref{correlation1}), quantum correlations $\langle \hat{A}\hat{B} \rangle_{\text{QM}}$, and experimental correlations $\langle \hat{A}\hat{B} \rangle_{\text{exp}}$. All of them are expressed in terms of probabilities for obtaining the results $\alpha$, $\beta$ when the SG magnets are oriented along $\boldsymbol{\hat{a}}$ and $\boldsymbol{\hat{b}}$, respectively:
\begin{equation}\label{corr-prob}
\langle \hat{A}\hat{B} \rangle_{K}=\sum_{\alpha,\beta}\alpha \beta P^{K}_{\boldsymbol{\hat{a}},\boldsymbol{\hat{b}}}(\alpha,\beta),
\end{equation}
with $K=\text{``exp''}$, $\text{``Bell''}$, and $\text{``QM''}$. These probabilities are given by
\begin{eqnarray}
P^{\text{exp}}_{\boldsymbol{\hat{a}},\boldsymbol{\hat{b}}}(\alpha,\beta)&=&\frac{N_{\boldsymbol{\hat{a}},\boldsymbol{\hat{b}}}(\alpha,\beta)}{N_{\text{tot}}}, \label{prob-exp}\\
 P_{\boldsymbol{\hat{a}},\boldsymbol{\hat{b}}}^{\text{Bell}}(\alpha,\beta)
 &=&\int p(\alpha|\boldsymbol{\hat{a}},\lambda)\,p(\beta|\boldsymbol{\hat{b}},\lambda)\rho(\lambda) d\lambda.\label{prob-bell}\\
  P_{\boldsymbol{\hat{a}},\boldsymbol{\hat{b}}}^{\text{QM}}(\alpha,\beta)&=&
  \left\langle\Pi^{A}(\boldsymbol{\hat{a}},\alpha)\otimes\Pi^{B}(\boldsymbol{\hat{b}},\beta)\right\rangle_{\Phi}, \label{prob-qm}
\end{eqnarray}
where $\Pi^{A}(\boldsymbol{\hat{a}},\alpha)=\frac{1}{2}\left(\sigma_0^{A}+ \alpha \boldsymbol{\hat{a}} \cdot \boldsymbol{\sigma}^{A}\right)$ and similarly for $\Pi^{B}(\boldsymbol{\hat{b}},\beta)$.

Equations (\ref{corr-prob}) and (\ref{prob-exp}) show how experimentalists construct $\langle \hat{A}\hat{B} \rangle_{\text{exp}}$. They keep separate record of, say, $N_{\boldsymbol{\hat{a}},\boldsymbol{\hat{b}}}(\alpha,\beta)$ and $N_{\boldsymbol{\hat{a}^{\prime}},\boldsymbol{\hat{b}}}(\alpha,\beta)$ (for all possible results $\alpha$ and $\beta$). This means that, e.g., Alice does not mix her measurement results $A_{\boldsymbol{\hat{a}}}^{\text{exp}}=+1$ and $A_{\boldsymbol{\hat{a}^{\prime}}}^{\text{exp}}=+1$, which correspond to having detected a particle at the ``upward'' detector in the two cases, but with her SG magnet oriented along $\boldsymbol{\hat{a}}$ in one case, and along $\boldsymbol{\hat{a}^{\prime}}$ in the other case. This is equivalent to having recorded $(+1,0)$ (for $\boldsymbol{\hat{a}}$) and 
$(0,+1)$ (for $\boldsymbol{\hat{a}^{\prime}}$). Hence, one should more properly write $A_{\boldsymbol{\hat{a}}}^{\text{exp}}=(+1,0)$ and $A_{\boldsymbol{\hat{a}^{\prime}}}^{\text{exp}}=(0,+1)$. This vectorial notation reflects what experimentalists actually do to calculate $\langle \hat{A}\hat{B} \rangle_{\text{exp}}$ from their raw data. 

On recalling the derivation of $|S|\leq 2$, in which use is made of $A_{\boldsymbol{\hat{a}}}=A_{\boldsymbol{\hat{a}^{\prime}}}=+1$, it becomes clear that experiments need not be constrained by Bell's inequality. This inequality rests on treating $A_{\boldsymbol{\hat{a}}}$ and $A_{\boldsymbol{\hat{a}^{\prime}}}$ as \emph{scalar} quantities, in contrast to what is effectively done with $A_{\boldsymbol{\hat{a}}}^{\text{exp}}$ and $A_{\boldsymbol{\hat{a}^{\prime}}}^{\text{exp}}$ in the actual experiments. 
On the other hand, Eqs.~(\ref{corr-prob}) and (\ref{prob-qm}) show that $\langle \hat{A}\hat{B} \rangle_{\text{QM}}$ has an intrinsic \emph{vectorial} nature and can therefore be in accordance with $\langle \hat{A}\hat{B} \rangle_{\text{exp}}$, as it is indeed the case. 

Experimental Bell tests are sometimes idealistically described as if they were an \emph{experimentum crucis}, i.e., as having the purpose of deciding which of two conflicting predictions is right, the one made by QM or the one made by local realism. This is not the actual case. In light of the above considerations, we may conclude that even an ideal, completely loophole-free Bell inequality violation would not imply that local realism is untenable.

\noindent \textbf{Acknowledgements}\\
Research partially funded by the European Commission, Grant Nr. H2020
MSCA RISE 2019 872049.

\end{document}